\documentstyle[12pt,prl,aps]{revtex}
\draft
\begin{document}
\title{Boltzmann equation approach for transport in systems subject to
microwave irradiation}
\author{Tobias Brandes}
\address{Department of Physics, Gakushuin University, 1--5--1
Mejiro, Toshima--ku, Tokyo 171, Japan}
\maketitle
\tighten
\vspace{0.5cm}
\begin{abstract}
We calculate the electronic transport properties of a system which is
irradiated
by a homogeneous microwave field. Within a Boltzmann equation approach, a
general expression for the conductivity tensor is derived and evaluated
for a quasi one-dimensional ballistic quantum wire and a two dimensional
system with impurity scattering. For the latter, deviations from the
Drude result are predicted for the conductivity tensor. This should be
observable in systems where the scattering rate depends noticeably on the
Fermi energy. The deviations are calculated explicitly for the case
of a 2d Silicon layer, where they strongly depend on the microwave
polarization and the Fermi energy.
\end{abstract}
\pacs{PACS: 72.10 Bg (General formulation of transport theory), 72.15 Lh
(Relaxation times and mean free paths), 72.30 +q
(High frequency effects; plasma effects)}
\section{Introduction}
Recently, frequency dependent effects have been discussed for the transport
properties of a large variety of systems.
Examples are periodically driven quantum wells \cite{Wag94},
photon--assisted tunneling through double barrier structures
\cite{Caietal90,IPT94}, and photo--induced transport through quantum
point contacts \cite{FH93}. In a two-dimensional system, the quantum Hall
effect has been investigated experimentally in the presence of an additional
microwave field \cite{Meietal92,Sunetal94}.

A common feature of these investigations is that the external frequency
dependent field (which  we call 'microwave' in the following) is an intrinsic
part of the system itself. This means that it does not
act as a 'probe field' to which the response is tested as in an ordinary AC
transport experiment. Rather, one has to think of transport in a DC 'probe'
field {\em in presence} of an additional, time dependent field that
eventually changes the DC transport properties as do other perturbations
like, e.g., the electron-electron or electron-phonon interaction. From the
theoretical point of view, one therefore has to reformulate the DC response
theory. In a quantum mechanical description, one would include the microwave
into the Hamiltonian which then becomes time dependent.

In this work, we use a different approach and derive the electronic
transport, namely the conductivity, in the semiclassical regime, where a
description in terms of the Boltzmann equation is expected to be valid.
Naturally, phenomena like localization, strong correlation effects or the
mesoscopic regime are beyond the validity of such a theory. As we show below,
however, even in the classical regime nontrivial effects can arise in the DC
transport due to the influence of an additional, time-dependent perturbation.
Moreover, while the transport in the linear regime always has to be
understood in the limit of a vanishing DC 'probe field', the microwave field
still could be arbitrarily strong and new, nonlinear effects can be expected.

In the following, we derive a general expression for the current linear in an
external 'probe field' in presence of an additional microwave within the
relaxation time approximation.
As an example, we first calculate the conductance of a clean
quantum wire, where the influence of the microwave, however, vanishes in the
DC limit. This is not the case for the second example of a two dimensional
silicon accumulation layer. There, the impurity scattering causes a
scattering rate that strongly depends on the electron's quasi momentum. The
latter is changed periodically due to the microwave which in turn eventually
gives a deviation of the conductivity from the standard Drude result. We
calculate this deviation to second order in the microwave and find it to be
negative for most values of the Fermi energy. Furthermore, it depends on the
microwave polarization and should be detectable in an ordinary DC measurement
of the 2d layer under microwave irradiation.

\section{Boltzmann equation}
The central concept for the description of electronic transport properties in
the framework of the Boltzmann equation is the semi--classical distribution
function $f({\bf r},{\bf k},t)$ in phase space.
Scattering of electron wave packets at impurities, phonons or other electrons
is governed by the equation (we set $\hbar=1$)
\begin{equation}
\label{boltzmannequation}
\left(\partial_{t}+{\bf v}_{{\bf k}}\partial_{{\bf r}}
+{\bf F}({\bf r},t)\partial_{{\bf k}} \right)f({\bf r},{\bf k},t)
=-{\cal L} f({\bf r},{\bf k},t),
\end{equation}
where ${\bf F}({\bf r},t)$ is the force exerted on the electrons by
an arbitrary electric field and ${\bf v}_{{\bf k}}={\bf k}/m$ the
velocity of electrons with bandmass $m$ and quasi momentum ${\bf k}$.
Here and in the following we do not include a magnetic field.
The collision term $-{\cal L} f({\bf r},{\bf k},t)$ is a nonlinear operator
acting on the distribution function $f$ and describes its change due to
scattering depending on the microscopic scattering mechanism.
Within the quasiclassical description of linear response quantities, we are
able to include exactly the influence of an arbitrary strong,
spatially  homogeneous microwave field.

The first step is to distinguish  in
Eq.~(\ref{boltzmannequation}) between two different contributions
to  ${\bf F}({\bf r},t)$: First, the
force $e{\bf E}_{0}(t)$ exerted on the electron by the microwave
( $e$ denotes the electron charge). Second, the
'probe field' ${\bf E}({\bf r},t)$ which is generated by driving a current
through the system. The proportionality between the latter and the probe
field is used to define linear response quantities like the conductivity.
Using this decomposition
${\bf F}({\bf r},t)=e{\bf E}_{0}(t)+e{\bf E}({\bf r},t)$,
the linear response to the probe field can be calculated.

We now solve the Boltzmann equation to linear order in the probe field
${\bf E}({\bf r},t)$. The distribution function is decomposed into
\begin{equation}
f({\bf r},{\bf k},t)=f_{0}({\bf k}(t))+\delta f({\bf r},{\bf k},t),
\end{equation}
where $f_{0}({\bf k}(t))$ is the solution in
absence of the probe field,
\begin{equation}
\label{f0solution}
f({\bf r},{\bf k},t)=f_{0}({\bf k}(t)),\qquad {\bf k}(t)={\bf k}-
e\int_{0}^t dt'\,{\bf E}_{0}(t')={\bf k}+\frac{e}{c} {\bf A}^{e}(t)
\end{equation}
where ${\bf A}^{e}(t)$ is the vector potential of the microwave field
with
$
{\bf E}_{0}(t)=-\partial {\bf A}^{e}(t)/\partial (ct).
$
(Note that the Fermi distribution function $f_{0}({\bf k})
= f({\bf r},{\bf k},t=0)$ gives zero upon
inserting into the collision term).
We obtain an equation for the change $\delta f({\bf r},{\bf k},t)$
due to the probe field
\begin{equation}
\label{linboltzmannequation}
\left(\partial_{t}+{\bf v}_{{\bf k}}\partial_{{\bf r}}
+e{\bf E}_{0}(t)\partial_{{\bf k}} \right)\delta f({\bf r},{\bf k},t)
+ e{\bf E}({\bf r},t)\partial_{{\bf k}} f_{0}({\bf k}(t))
=-{\cal L} \left[f_{0}({\bf k}(t))+\delta f({\bf r},{\bf k},t)\right].
\end{equation}
Note that in Eq.~(\ref{linboltzmannequation}) the microwave field ${\bf
E}_{0}(t)$ is still fully included whereas the probe field ${\bf E}({\bf
r},t)$ is considered only to lowest order.

In general, since the Boltzmann equation is a nonlinear integral equation,
it can not be solved exactly.
In the following, we restrict ourselves to
the relaxation time approximation,
$
{\cal L} \delta f({\bf r},{\bf k},t)=\tau ^{-1}({\bf k})\delta f({\bf r},{\bf
k},t),
$
where
$\tau ^{-1}({\bf k})$ is the scattering rate for electrons with
quasi momentum ${\bf k}$.
In the case of elastic impurity scattering,
$
\tau ^{-1}({\bf k})=1/L^d\sum_{{\bf k'}}W_{{\bf k k'}},
$
where
$
W_{{\bf k k'}}=2\pi n_{i}\delta (\varepsilon _{k}-
\varepsilon _{k'}) |\langle{\bf k}|V|{\bf k'}\rangle|^2
$
and $V$ is the potential of impurities with a concentration
$n_{i}$ in a system of volume $L^d $.
In the limit of weak fields ${\bf F}({\bf r},t)\to 0$, the relaxation time
approximation becomes
exact for isotropic potentials such that $W_{{\bf k
k'}}=W_{{\bf |k| |k'|}}$.
In general, this approximation replaces
the effect of the collision term by a rate $\tau ^{-
1}({\bf k})$ describing the relaxation of a deviation $\delta f$ of the
distribution function towards its quasi-equilibrium value $f_{0}({\bf k}(t))$,
Eq.~(\ref{f0solution}) due to scattering.

It is a standard procedure to solve the resulting
first order partial differential equation.
The solution can be obtained by the methods of characteristics \cite{Courant}
and reads
\begin{eqnarray}
\label{boltzsolution}
\delta f({\bf q},{\bf k},t)&=&
-\partial_{{\bf k}} f_{0}\left({\bf k}(t)\right)
\int_{0}^t dt'\,e{\bf E}({\bf q},t')
\exp \left\{ -\int_{t'}^tds\, g\left[
{\bf k}(t)+e\int_{0}^s dt''\,{\bf E}_{0}(t'')   \right]\right\}\nonumber\\
g({\bf k})&:=&\tau^{-1}({\bf k})+i{\bf v}_{{\bf k}}{\bf q},
\end{eqnarray}
where we Fourier transformed ${\bf r}\to {\bf q}$.
The current density is obtained from the resulting
change $\delta f({\bf q},{\bf k},t)$ of the distribution function as
\begin{eqnarray}
\label{currentformula}
{\bf j}({\bf q},t)&\equiv&\frac{e}{L^{d}}\sum_{{\bf k}}{\bf v}_{{\bf k}}
 f({\bf q},{\bf k},t)=\frac{e^{2}}{L^{d}m}\sum_{{\bf k}}
\left[{\bf k}-\frac{e}{c}{\bf A}^{e}(t)\right]
\left(-\partial_{{\bf k}} f_{0}({\bf k})\right) \times\nonumber\\
&\times&
\int_{0}^t dt'\,{\bf E}({\bf q},t')
\exp \left\{ -\int_{t'}^tds\, g\left[
{\bf k}+e\int_{0}^s dt''\,{\bf E}_{0}(t'')   \right]\right\}
-\frac{e^{2}}{mc}n_{e}{\bf A}^{e}(t).
\end{eqnarray}
Here, we shifted the ${\bf k}$-summation according to ${\bf k}\to {\bf k}-(e/c)
{\bf A}^{e}(t)$ in Eq.~(\ref{currentformula}), and $n_{e}$ is the electronic
density.

Eq.~(\ref{currentformula}) is the central result, expressing the response of
the current to the probe field, and valid within the relaxation time
approximation for arbitrary microwave field strength ${\bf E_{0}}$.
The scattering rate $\tau^{-1}({\bf k})$, however, in general has a
nontrivial dependence on ${\bf k}$ which makes it impossible to evaluate
Eq.~(\ref{currentformula}) analytically.

The conductivity $\sigma ({\bf q},z)$ is defined by the component of the
current density ${\bf j}_{pr}({\bf q},t)$ with a time dependence
proportional to $\exp (-izt)$, the time dependence of the probe field:
\begin{equation}
\label{jpr}
{\bf j}_{pr}({\bf q},t)=\sigma ({\bf q},z)\times {\bf E}({\bf q})\exp(-izt).
\end{equation}
In the case of a constant scattering rate
$
\tau^{-1} ({\bf k})=\tau ^{-1},
$
the $t'$ integration in Eq.~(\ref{currentformula}) can be performed and
one obtains an analytical expression for the conductivity
Eq.~(\ref{jpr}).
We evaluate its real part $\sigma '$
for a quasi one dimensional wire,
where it serves as a definition of the {\em
conductance} \cite{FL81,KM89}. It can be evaluated for the
ballistic case, which is defined by the limit of vanishing scattering rate,
$\tau ^{-1}\to 0$,
describing a 'clean' quantum wire in the limit of only one occupied subband
in absence of any scattering. The result is
\begin{equation}
\label{sigmansum}
\sigma '(q,\omega )=\sum_{n=-
\infty}^{\infty}J_{n}^{2}\left(\frac{eE_{0}q}{m\Omega ^{2}}\right)
\sigma _{0}'(q,\omega -n\Omega )\frac{\omega }{\omega -n\Omega },
\end{equation}
where
$
\sigma '_{0}(q,\omega )=(e^{2}/L)
\sum_{k} v_{k}\left(-\partial_{k}f_{0}(k)\right)
\pi \delta (\omega -v_{k}q)
$
is the real part of the conductivity in absence of the microwave.
This results coincides with a quantum mechanical calculation \cite{BKP95}
for $\hbar\Omega \ll \varepsilon _{F}$ ($\varepsilon _{F}$ Fermi energy);
the appearance of the Bessel functions is similar to the Tien--Gordon theory
\cite{TG63} of photo-assisted tunneling in superconducting junctions.
Thus, we have expressed the conductivity in {\em presence} of the microwave in
terms of the conductivity $\sigma '_{0}(q,\omega )$ {\em in absence} of the
microwave. The term $n=0$ in Eq.~(\ref{sigmansum}) just yields $\sigma
'_{0}(q,\omega )$, while the terms $n\ne 0$ describe the influence of the
microwave. The latter, however, vanishes in the limit of zero
(probe field) frequency
$\omega \to 0$. In particular, this means that the {\em DC-conductance} of a
clean quantum wire is {\em not} changed under the influence of a homogeneous
microwave  field \cite{BKP95}. A non-vanishing effect can be expected only in
presence of an additional scattering mechanism like impurities or a potential
barrier.

\section{Two dimensional system}
As a further application of our formalism, we evaluated the conductivity
tensor $\sigma ({\bf q}=0,z)$ for a two--dimensional system from
Eq.~(\ref{jpr}) and Eq.~(\ref{currentformula}) to second order in the
microwave field ${\bf E_{0}}$. The electron gas and the microwave
polarization ${\bf E}_{0}=E_{0}{\bf e}_{x}$ are assumed to lie in the $x$-$y$
plane. Here, we give the zero-temperature result which holds for
$k_{B}T\ll\varepsilon _{F}$, although corresponding
expressions for $T>0$ can easily be derived. We obtain
\begin{eqnarray}
\label{finalsecond}
\sigma _{xx}&=:& \sigma _{0}+ \Delta \sigma _{xx}=
\sigma _{0}\left\{1+\frac{(\Omega
\tau(\varepsilon _{F}) )^{2}}{1+(\Omega \tau(\varepsilon _{F})
)^{2}}\left[\frac{3}{8}
\left(\frac{v_{F}eE_{0}}{\Omega ^{2}}(\tau ^{-1})'(\varepsilon _{F})\right)^{2}
-\frac{e^{2}E_{0}^{2}}{2m\Omega ^{4}\tau(\varepsilon _{F}) }
(\tau ^{-1})'(\varepsilon _{F})
\right]\right.\nonumber\\
&-&\left.\left(\frac{eE_{0}}{\Omega }\right)^{2}\frac{\tau (\varepsilon
_{F})}{4m}
\left[ (\tau ^{-1})'(\varepsilon _{F}) +\frac{3}{2}\varepsilon _{F}
(\tau ^{-1})''(\varepsilon _{F})\right]\right\} +O(E_{0})^{4}\nonumber\\
\sigma _{yy}&=:& \sigma _{0}+ \Delta \sigma _{yy}=
\sigma _{0}\left\{1+\frac{(\Omega
\tau(\varepsilon _{F}) )^{2}}{1+(\Omega \tau(\varepsilon _{F})
)^{2}}\frac{1}{8}
\left(\frac{v_{F}eE_{0}}{\Omega ^{2}}(\tau ^{-1})'(\varepsilon _{F})\right)^{2}
\right.\nonumber\\
&-&\left.\left(\frac{eE_{0}}{\Omega }\right)^{2}\frac{\tau (\varepsilon
_{F})}{4m}
\left[ (\tau ^{-1})'(\varepsilon _{F}) +\frac{1}{2}\varepsilon _{F}
(\tau ^{-1})''(\varepsilon _{F})\right]\right\} +O(E_{0})^{4},
\end{eqnarray}
and $\sigma _{xy}=\sigma _{yx}=0$.
Here, $\sigma _{0}=e^{2}n_{s}\tau(\varepsilon _{F})/m$ is the zero frequency
(Drude) conductivity in absence of the microwave ($m$ denotes the electron
band mass).
Within the relaxation time approximation, Eq.~(\ref{finalsecond}) is
valid for both elastic an inelastic scattering, described by the rate
$\tau (\varepsilon _{F})^{-1}$.
Due to the appearance of its derivatives in Eq.~(\ref{finalsecond}), one
needs to know the dependence of $\tau^{-1}$ on the Fermi energy $\varepsilon
_{F}$.

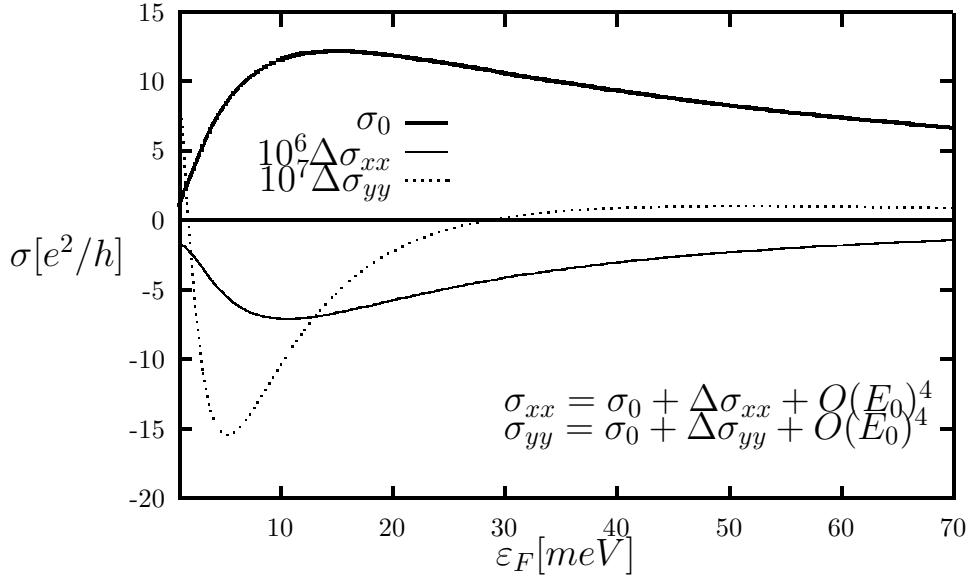
\begin{figure}[]
\unitlength1cm
\begin{picture}(13,10)
\large
\put(0,.5){
\setlength{\unitlength}{0.240900pt}
\ifx\plotpoint\undefined\newsavebox{\plotpoint}\fi
\sbox{\plotpoint}{\rule[-0.500pt]{1.000pt}{1.000pt}}%
\begin{picture}(1500,900)(0,0)
\font\gnuplot=cmr10 at 10pt
\gnuplot
\sbox{\plotpoint}{\rule[-0.500pt]{1.000pt}{1.000pt}}%
\put(220.0,550.0){\rule[-0.500pt]{292.934pt}{1.000pt}}
\put(220.0,113.0){\rule[-0.500pt]{4.818pt}{1.000pt}}
\put(198,113){\makebox(0,0)[r]{-20}}
\put(1416.0,113.0){\rule[-0.500pt]{4.818pt}{1.000pt}}
\put(220.0,222.0){\rule[-0.500pt]{4.818pt}{1.000pt}}
\put(198,222){\makebox(0,0)[r]{-15}}
\put(1416.0,222.0){\rule[-0.500pt]{4.818pt}{1.000pt}}
\put(220.0,331.0){\rule[-0.500pt]{4.818pt}{1.000pt}}
\put(198,331){\makebox(0,0)[r]{-10}}
\put(1416.0,331.0){\rule[-0.500pt]{4.818pt}{1.000pt}}
\put(220.0,440.0){\rule[-0.500pt]{4.818pt}{1.000pt}}
\put(198,440){\makebox(0,0)[r]{-5}}
\put(1416.0,440.0){\rule[-0.500pt]{4.818pt}{1.000pt}}
\put(220.0,550.0){\rule[-0.500pt]{4.818pt}{1.000pt}}
\put(198,550){\makebox(0,0)[r]{0}}
\put(1416.0,550.0){\rule[-0.500pt]{4.818pt}{1.000pt}}
\put(220.0,659.0){\rule[-0.500pt]{4.818pt}{1.000pt}}
\put(198,659){\makebox(0,0)[r]{5}}
\put(1416.0,659.0){\rule[-0.500pt]{4.818pt}{1.000pt}}
\put(220.0,768.0){\rule[-0.500pt]{4.818pt}{1.000pt}}
\put(198,768){\makebox(0,0)[r]{10}}
\put(1416.0,768.0){\rule[-0.500pt]{4.818pt}{1.000pt}}
\put(220.0,877.0){\rule[-0.500pt]{4.818pt}{1.000pt}}
\put(198,877){\makebox(0,0)[r]{15}}
\put(1416.0,877.0){\rule[-0.500pt]{4.818pt}{1.000pt}}
\put(379.0,113.0){\rule[-0.500pt]{1.000pt}{4.818pt}}
\put(379,68){\makebox(0,0){10}}
\put(379.0,857.0){\rule[-0.500pt]{1.000pt}{4.818pt}}
\put(555.0,113.0){\rule[-0.500pt]{1.000pt}{4.818pt}}
\put(555,68){\makebox(0,0){20}}
\put(555.0,857.0){\rule[-0.500pt]{1.000pt}{4.818pt}}
\put(731.0,113.0){\rule[-0.500pt]{1.000pt}{4.818pt}}
\put(731,68){\makebox(0,0){30}}
\put(731.0,857.0){\rule[-0.500pt]{1.000pt}{4.818pt}}
\put(907.0,113.0){\rule[-0.500pt]{1.000pt}{4.818pt}}
\put(907,68){\makebox(0,0){40}}
\put(907.0,857.0){\rule[-0.500pt]{1.000pt}{4.818pt}}
\put(1084.0,113.0){\rule[-0.500pt]{1.000pt}{4.818pt}}
\put(1084,68){\makebox(0,0){50}}
\put(1084.0,857.0){\rule[-0.500pt]{1.000pt}{4.818pt}}
\put(1260.0,113.0){\rule[-0.500pt]{1.000pt}{4.818pt}}
\put(1260,68){\makebox(0,0){60}}
\put(1260.0,857.0){\rule[-0.500pt]{1.000pt}{4.818pt}}
\put(1436.0,113.0){\rule[-0.500pt]{1.000pt}{4.818pt}}
\put(1436,68){\makebox(0,0){70}}
\put(1436.0,857.0){\rule[-0.500pt]{1.000pt}{4.818pt}}
\put(220.0,113.0){\rule[-0.500pt]{292.934pt}{1.000pt}}
\put(1436.0,113.0){\rule[-0.500pt]{1.000pt}{184.048pt}}
\put(220.0,877.0){\rule[-0.500pt]{292.934pt}{1.000pt}}
\put(45,495){\makebox(0,0){$\sigma [e^2/h]$}}
\put(828,23){\makebox(0,0){$\varepsilon_{F}[meV]$}}
\put(731,266){\makebox(0,0)[l]{$\sigma_{xx}=\sigma_0+\Delta
\sigma_{xx}+O(E_0)^4$}}
\put(731,222){\makebox(0,0)[l]{$\sigma_{yy}=\sigma_0+\Delta
\sigma_{yy}+O(E_0)^4$}}
\put(220.0,113.0){\rule[-0.500pt]{1.000pt}{184.048pt}}
\put(555,702){\makebox(0,0)[r]{$\sigma_0$}}
\put(577.0,702.0){\rule[-0.500pt]{15.899pt}{1.000pt}}
\put(220,571){\usebox{\plotpoint}}
\multiput(221.83,571.00)(0.491,1.283){16}{\rule{0.118pt}{2.833pt}}
\multiput(217.92,571.00)(12.000,25.119){2}{\rule{1.000pt}{1.417pt}}
\multiput(233.83,602.00)(0.492,1.220){18}{\rule{0.118pt}{2.712pt}}
\multiput(229.92,602.00)(13.000,26.372){2}{\rule{1.000pt}{1.356pt}}
\multiput(246.83,634.00)(0.491,1.239){16}{\rule{0.118pt}{2.750pt}}
\multiput(242.92,634.00)(12.000,24.292){2}{\rule{1.000pt}{1.375pt}}
\multiput(258.83,664.00)(0.491,1.108){16}{\rule{0.118pt}{2.500pt}}
\multiput(254.92,664.00)(12.000,21.811){2}{\rule{1.000pt}{1.250pt}}
\multiput(270.83,691.00)(0.491,0.890){16}{\rule{0.118pt}{2.083pt}}
\multiput(266.92,691.00)(12.000,17.676){2}{\rule{1.000pt}{1.042pt}}
\multiput(282.83,713.00)(0.492,0.740){18}{\rule{0.118pt}{1.788pt}}
\multiput(278.92,713.00)(13.000,16.288){2}{\rule{1.000pt}{0.894pt}}
\multiput(295.83,733.00)(0.491,0.628){16}{\rule{0.118pt}{1.583pt}}
\multiput(291.92,733.00)(12.000,12.714){2}{\rule{1.000pt}{0.792pt}}
\multiput(307.83,749.00)(0.491,0.498){16}{\rule{0.118pt}{1.333pt}}
\multiput(303.92,749.00)(12.000,10.233){2}{\rule{1.000pt}{0.667pt}}
\multiput(318.00,763.83)(0.498,0.491){16}{\rule{1.333pt}{0.118pt}}
\multiput(318.00,759.92)(10.233,12.000){2}{\rule{0.667pt}{1.000pt}}
\multiput(331.00,775.83)(0.603,0.485){10}{\rule{1.583pt}{0.117pt}}
\multiput(331.00,771.92)(8.714,9.000){2}{\rule{0.792pt}{1.000pt}}
\multiput(343.00,784.83)(0.677,0.481){8}{\rule{1.750pt}{0.116pt}}
\multiput(343.00,780.92)(8.368,8.000){2}{\rule{0.875pt}{1.000pt}}
\multiput(355.00,792.84)(0.887,0.462){4}{\rule{2.250pt}{0.111pt}}
\multiput(355.00,788.92)(7.330,6.000){2}{\rule{1.125pt}{1.000pt}}
\multiput(367.00,798.86)(1.169,0.424){2}{\rule{2.850pt}{0.102pt}}
\multiput(367.00,794.92)(7.085,5.000){2}{\rule{1.425pt}{1.000pt}}
\put(380,801.92){\rule{2.891pt}{1.000pt}}
\multiput(380.00,799.92)(6.000,4.000){2}{\rule{1.445pt}{1.000pt}}
\put(392,805.42){\rule{2.891pt}{1.000pt}}
\multiput(392.00,803.92)(6.000,3.000){2}{\rule{1.445pt}{1.000pt}}
\put(404,808.42){\rule{3.132pt}{1.000pt}}
\multiput(404.00,806.92)(6.500,3.000){2}{\rule{1.566pt}{1.000pt}}
\put(417,810.42){\rule{2.891pt}{1.000pt}}
\multiput(417.00,809.92)(6.000,1.000){2}{\rule{1.445pt}{1.000pt}}
\put(429,811.42){\rule{2.891pt}{1.000pt}}
\multiput(429.00,810.92)(6.000,1.000){2}{\rule{1.445pt}{1.000pt}}
\put(441,812.42){\rule{2.891pt}{1.000pt}}
\multiput(441.00,811.92)(6.000,1.000){2}{\rule{1.445pt}{1.000pt}}
\put(490,812.42){\rule{3.132pt}{1.000pt}}
\multiput(490.00,812.92)(6.500,-1.000){2}{\rule{1.566pt}{1.000pt}}
\put(503,811.42){\rule{2.891pt}{1.000pt}}
\multiput(503.00,811.92)(6.000,-1.000){2}{\rule{1.445pt}{1.000pt}}
\put(515,810.42){\rule{2.891pt}{1.000pt}}
\multiput(515.00,810.92)(6.000,-1.000){2}{\rule{1.445pt}{1.000pt}}
\put(527,809.42){\rule{2.891pt}{1.000pt}}
\multiput(527.00,809.92)(6.000,-1.000){2}{\rule{1.445pt}{1.000pt}}
\put(539,807.92){\rule{3.132pt}{1.000pt}}
\multiput(539.00,808.92)(6.500,-2.000){2}{\rule{1.566pt}{1.000pt}}
\put(552,806.42){\rule{2.891pt}{1.000pt}}
\multiput(552.00,806.92)(6.000,-1.000){2}{\rule{1.445pt}{1.000pt}}
\put(564,804.92){\rule{2.891pt}{1.000pt}}
\multiput(564.00,805.92)(6.000,-2.000){2}{\rule{1.445pt}{1.000pt}}
\put(576,802.92){\rule{2.891pt}{1.000pt}}
\multiput(576.00,803.92)(6.000,-2.000){2}{\rule{1.445pt}{1.000pt}}
\put(588,800.92){\rule{3.132pt}{1.000pt}}
\multiput(588.00,801.92)(6.500,-2.000){2}{\rule{1.566pt}{1.000pt}}
\put(601,799.42){\rule{2.891pt}{1.000pt}}
\multiput(601.00,799.92)(6.000,-1.000){2}{\rule{1.445pt}{1.000pt}}
\put(613,797.92){\rule{2.891pt}{1.000pt}}
\multiput(613.00,798.92)(6.000,-2.000){2}{\rule{1.445pt}{1.000pt}}
\put(625,795.92){\rule{3.132pt}{1.000pt}}
\multiput(625.00,796.92)(6.500,-2.000){2}{\rule{1.566pt}{1.000pt}}
\put(638,793.92){\rule{2.891pt}{1.000pt}}
\multiput(638.00,794.92)(6.000,-2.000){2}{\rule{1.445pt}{1.000pt}}
\put(650,791.92){\rule{2.891pt}{1.000pt}}
\multiput(650.00,792.92)(6.000,-2.000){2}{\rule{1.445pt}{1.000pt}}
\put(662,789.92){\rule{2.891pt}{1.000pt}}
\multiput(662.00,790.92)(6.000,-2.000){2}{\rule{1.445pt}{1.000pt}}
\put(674,787.92){\rule{3.132pt}{1.000pt}}
\multiput(674.00,788.92)(6.500,-2.000){2}{\rule{1.566pt}{1.000pt}}
\put(687,785.92){\rule{2.891pt}{1.000pt}}
\multiput(687.00,786.92)(6.000,-2.000){2}{\rule{1.445pt}{1.000pt}}
\put(699,783.92){\rule{2.891pt}{1.000pt}}
\multiput(699.00,784.92)(6.000,-2.000){2}{\rule{1.445pt}{1.000pt}}
\put(711,781.42){\rule{3.132pt}{1.000pt}}
\multiput(711.00,782.92)(6.500,-3.000){2}{\rule{1.566pt}{1.000pt}}
\put(724,778.92){\rule{2.891pt}{1.000pt}}
\multiput(724.00,779.92)(6.000,-2.000){2}{\rule{1.445pt}{1.000pt}}
\put(736,776.92){\rule{2.891pt}{1.000pt}}
\multiput(736.00,777.92)(6.000,-2.000){2}{\rule{1.445pt}{1.000pt}}
\put(748,774.92){\rule{2.891pt}{1.000pt}}
\multiput(748.00,775.92)(6.000,-2.000){2}{\rule{1.445pt}{1.000pt}}
\put(760,772.92){\rule{3.132pt}{1.000pt}}
\multiput(760.00,773.92)(6.500,-2.000){2}{\rule{1.566pt}{1.000pt}}
\put(773,770.92){\rule{2.891pt}{1.000pt}}
\multiput(773.00,771.92)(6.000,-2.000){2}{\rule{1.445pt}{1.000pt}}
\put(785,768.92){\rule{2.891pt}{1.000pt}}
\multiput(785.00,769.92)(6.000,-2.000){2}{\rule{1.445pt}{1.000pt}}
\put(797,766.92){\rule{3.132pt}{1.000pt}}
\multiput(797.00,767.92)(6.500,-2.000){2}{\rule{1.566pt}{1.000pt}}
\put(810,765.42){\rule{2.891pt}{1.000pt}}
\multiput(810.00,765.92)(6.000,-1.000){2}{\rule{1.445pt}{1.000pt}}
\put(822,763.92){\rule{2.891pt}{1.000pt}}
\multiput(822.00,764.92)(6.000,-2.000){2}{\rule{1.445pt}{1.000pt}}
\put(834,761.92){\rule{2.891pt}{1.000pt}}
\multiput(834.00,762.92)(6.000,-2.000){2}{\rule{1.445pt}{1.000pt}}
\put(846,759.92){\rule{3.132pt}{1.000pt}}
\multiput(846.00,760.92)(6.500,-2.000){2}{\rule{1.566pt}{1.000pt}}
\put(859,757.92){\rule{2.891pt}{1.000pt}}
\multiput(859.00,758.92)(6.000,-2.000){2}{\rule{1.445pt}{1.000pt}}
\put(871,755.92){\rule{2.891pt}{1.000pt}}
\multiput(871.00,756.92)(6.000,-2.000){2}{\rule{1.445pt}{1.000pt}}
\put(883,753.92){\rule{3.132pt}{1.000pt}}
\multiput(883.00,754.92)(6.500,-2.000){2}{\rule{1.566pt}{1.000pt}}
\put(896,752.42){\rule{2.891pt}{1.000pt}}
\multiput(896.00,752.92)(6.000,-1.000){2}{\rule{1.445pt}{1.000pt}}
\put(908,750.92){\rule{2.891pt}{1.000pt}}
\multiput(908.00,751.92)(6.000,-2.000){2}{\rule{1.445pt}{1.000pt}}
\put(920,748.92){\rule{2.891pt}{1.000pt}}
\multiput(920.00,749.92)(6.000,-2.000){2}{\rule{1.445pt}{1.000pt}}
\put(932,746.92){\rule{3.132pt}{1.000pt}}
\multiput(932.00,747.92)(6.500,-2.000){2}{\rule{1.566pt}{1.000pt}}
\put(945,745.42){\rule{2.891pt}{1.000pt}}
\multiput(945.00,745.92)(6.000,-1.000){2}{\rule{1.445pt}{1.000pt}}
\put(957,743.92){\rule{2.891pt}{1.000pt}}
\multiput(957.00,744.92)(6.000,-2.000){2}{\rule{1.445pt}{1.000pt}}
\put(969,741.92){\rule{3.132pt}{1.000pt}}
\multiput(969.00,742.92)(6.500,-2.000){2}{\rule{1.566pt}{1.000pt}}
\put(982,740.42){\rule{2.891pt}{1.000pt}}
\multiput(982.00,740.92)(6.000,-1.000){2}{\rule{1.445pt}{1.000pt}}
\put(994,738.92){\rule{2.891pt}{1.000pt}}
\multiput(994.00,739.92)(6.000,-2.000){2}{\rule{1.445pt}{1.000pt}}
\put(1006,736.92){\rule{2.891pt}{1.000pt}}
\multiput(1006.00,737.92)(6.000,-2.000){2}{\rule{1.445pt}{1.000pt}}
\put(1018,735.42){\rule{3.132pt}{1.000pt}}
\multiput(1018.00,735.92)(6.500,-1.000){2}{\rule{1.566pt}{1.000pt}}
\put(1031,733.92){\rule{2.891pt}{1.000pt}}
\multiput(1031.00,734.92)(6.000,-2.000){2}{\rule{1.445pt}{1.000pt}}
\put(1043,732.42){\rule{2.891pt}{1.000pt}}
\multiput(1043.00,732.92)(6.000,-1.000){2}{\rule{1.445pt}{1.000pt}}
\put(1055,730.92){\rule{3.132pt}{1.000pt}}
\multiput(1055.00,731.92)(6.500,-2.000){2}{\rule{1.566pt}{1.000pt}}
\put(1068,729.42){\rule{2.891pt}{1.000pt}}
\multiput(1068.00,729.92)(6.000,-1.000){2}{\rule{1.445pt}{1.000pt}}
\put(1080,727.92){\rule{2.891pt}{1.000pt}}
\multiput(1080.00,728.92)(6.000,-2.000){2}{\rule{1.445pt}{1.000pt}}
\put(1092,726.42){\rule{2.891pt}{1.000pt}}
\multiput(1092.00,726.92)(6.000,-1.000){2}{\rule{1.445pt}{1.000pt}}
\put(1104,724.92){\rule{3.132pt}{1.000pt}}
\multiput(1104.00,725.92)(6.500,-2.000){2}{\rule{1.566pt}{1.000pt}}
\put(1117,723.42){\rule{2.891pt}{1.000pt}}
\multiput(1117.00,723.92)(6.000,-1.000){2}{\rule{1.445pt}{1.000pt}}
\put(1129,722.42){\rule{2.891pt}{1.000pt}}
\multiput(1129.00,722.92)(6.000,-1.000){2}{\rule{1.445pt}{1.000pt}}
\put(1141,720.92){\rule{2.891pt}{1.000pt}}
\multiput(1141.00,721.92)(6.000,-2.000){2}{\rule{1.445pt}{1.000pt}}
\put(1153,719.42){\rule{3.132pt}{1.000pt}}
\multiput(1153.00,719.92)(6.500,-1.000){2}{\rule{1.566pt}{1.000pt}}
\put(1166,717.92){\rule{2.891pt}{1.000pt}}
\multiput(1166.00,718.92)(6.000,-2.000){2}{\rule{1.445pt}{1.000pt}}
\put(1178,716.42){\rule{2.891pt}{1.000pt}}
\multiput(1178.00,716.92)(6.000,-1.000){2}{\rule{1.445pt}{1.000pt}}
\put(1190,715.42){\rule{3.132pt}{1.000pt}}
\multiput(1190.00,715.92)(6.500,-1.000){2}{\rule{1.566pt}{1.000pt}}
\put(1203,714.42){\rule{2.891pt}{1.000pt}}
\multiput(1203.00,714.92)(6.000,-1.000){2}{\rule{1.445pt}{1.000pt}}
\put(1215,712.92){\rule{2.891pt}{1.000pt}}
\multiput(1215.00,713.92)(6.000,-2.000){2}{\rule{1.445pt}{1.000pt}}
\put(1227,711.42){\rule{2.891pt}{1.000pt}}
\multiput(1227.00,711.92)(6.000,-1.000){2}{\rule{1.445pt}{1.000pt}}
\put(1239,710.42){\rule{3.132pt}{1.000pt}}
\multiput(1239.00,710.92)(6.500,-1.000){2}{\rule{1.566pt}{1.000pt}}
\put(1252,709.42){\rule{2.891pt}{1.000pt}}
\multiput(1252.00,709.92)(6.000,-1.000){2}{\rule{1.445pt}{1.000pt}}
\put(1264,707.92){\rule{2.891pt}{1.000pt}}
\multiput(1264.00,708.92)(6.000,-2.000){2}{\rule{1.445pt}{1.000pt}}
\put(1276,706.42){\rule{3.132pt}{1.000pt}}
\multiput(1276.00,706.92)(6.500,-1.000){2}{\rule{1.566pt}{1.000pt}}
\put(1289,705.42){\rule{2.891pt}{1.000pt}}
\multiput(1289.00,705.92)(6.000,-1.000){2}{\rule{1.445pt}{1.000pt}}
\put(1301,704.42){\rule{2.891pt}{1.000pt}}
\multiput(1301.00,704.92)(6.000,-1.000){2}{\rule{1.445pt}{1.000pt}}
\put(1313,703.42){\rule{2.891pt}{1.000pt}}
\multiput(1313.00,703.92)(6.000,-1.000){2}{\rule{1.445pt}{1.000pt}}
\put(1325,702.42){\rule{3.132pt}{1.000pt}}
\multiput(1325.00,702.92)(6.500,-1.000){2}{\rule{1.566pt}{1.000pt}}
\put(1338,701.42){\rule{2.891pt}{1.000pt}}
\multiput(1338.00,701.92)(6.000,-1.000){2}{\rule{1.445pt}{1.000pt}}
\put(1350,700.42){\rule{2.891pt}{1.000pt}}
\multiput(1350.00,700.92)(6.000,-1.000){2}{\rule{1.445pt}{1.000pt}}
\put(1362,698.92){\rule{3.132pt}{1.000pt}}
\multiput(1362.00,699.92)(6.500,-2.000){2}{\rule{1.566pt}{1.000pt}}
\put(1375,697.42){\rule{2.891pt}{1.000pt}}
\multiput(1375.00,697.92)(6.000,-1.000){2}{\rule{1.445pt}{1.000pt}}
\put(1387,696.42){\rule{2.891pt}{1.000pt}}
\multiput(1387.00,696.92)(6.000,-1.000){2}{\rule{1.445pt}{1.000pt}}
\put(1399,695.42){\rule{2.891pt}{1.000pt}}
\multiput(1399.00,695.92)(6.000,-1.000){2}{\rule{1.445pt}{1.000pt}}
\put(1411,694.42){\rule{3.132pt}{1.000pt}}
\multiput(1411.00,694.92)(6.500,-1.000){2}{\rule{1.566pt}{1.000pt}}
\put(1424,693.42){\rule{2.891pt}{1.000pt}}
\multiput(1424.00,693.92)(6.000,-1.000){2}{\rule{1.445pt}{1.000pt}}
\put(453.0,815.0){\rule[-0.500pt]{8.913pt}{1.000pt}}
\sbox{\plotpoint}{\rule[-0.175pt]{0.350pt}{0.350pt}}%
\put(555,657){\makebox(0,0)[r]{$10^6 \Delta \sigma_{xx}$}}
\put(577.0,657.0){\rule[-0.175pt]{15.899pt}{0.350pt}}
\put(220,514){\usebox{\plotpoint}}
\multiput(220.00,513.02)(0.788,-0.504){13}{\rule{0.612pt}{0.121pt}}
\multiput(220.00,513.27)(10.729,-8.000){2}{\rule{0.306pt}{0.350pt}}
\multiput(232.48,504.07)(0.502,-0.549){23}{\rule{0.121pt}{0.464pt}}
\multiput(231.27,505.04)(13.000,-13.036){2}{\rule{0.350pt}{0.232pt}}
\multiput(245.48,489.70)(0.502,-0.685){21}{\rule{0.121pt}{0.554pt}}
\multiput(244.27,490.85)(12.000,-14.850){2}{\rule{0.350pt}{0.277pt}}
\multiput(257.48,473.58)(0.502,-0.729){21}{\rule{0.121pt}{0.583pt}}
\multiput(256.27,474.79)(12.000,-15.789){2}{\rule{0.350pt}{0.292pt}}
\multiput(269.48,456.82)(0.502,-0.641){21}{\rule{0.121pt}{0.525pt}}
\multiput(268.27,457.91)(12.000,-13.910){2}{\rule{0.350pt}{0.262pt}}
\multiput(281.00,443.02)(0.508,-0.502){23}{\rule{0.438pt}{0.121pt}}
\multiput(281.00,443.27)(12.092,-13.000){2}{\rule{0.219pt}{0.350pt}}
\multiput(294.00,430.02)(0.558,-0.502){19}{\rule{0.469pt}{0.121pt}}
\multiput(294.00,430.27)(11.026,-11.000){2}{\rule{0.235pt}{0.350pt}}
\multiput(306.00,419.02)(0.788,-0.504){13}{\rule{0.612pt}{0.121pt}}
\multiput(306.00,419.27)(10.729,-8.000){2}{\rule{0.306pt}{0.350pt}}
\multiput(318.00,411.02)(0.994,-0.504){11}{\rule{0.737pt}{0.121pt}}
\multiput(318.00,411.27)(11.469,-7.000){2}{\rule{0.369pt}{0.350pt}}
\multiput(331.00,404.02)(1.820,-0.509){5}{\rule{1.137pt}{0.123pt}}
\multiput(331.00,404.27)(9.639,-4.000){2}{\rule{0.569pt}{0.350pt}}
\multiput(343.00,400.02)(2.913,-0.516){3}{\rule{1.487pt}{0.124pt}}
\multiput(343.00,400.27)(8.913,-3.000){2}{\rule{0.744pt}{0.350pt}}
\put(355,396.27){\rule{2.188pt}{0.350pt}}
\multiput(355.00,397.27)(7.460,-2.000){2}{\rule{1.094pt}{0.350pt}}
\put(367,394.77){\rule{3.132pt}{0.350pt}}
\multiput(367.00,395.27)(6.500,-1.000){2}{\rule{1.566pt}{0.350pt}}
\put(380,393.77){\rule{2.891pt}{0.350pt}}
\multiput(380.00,394.27)(6.000,-1.000){2}{\rule{1.445pt}{0.350pt}}
\put(392,393.77){\rule{2.891pt}{0.350pt}}
\multiput(392.00,393.27)(6.000,1.000){2}{\rule{1.445pt}{0.350pt}}
\put(404,394.77){\rule{3.132pt}{0.350pt}}
\multiput(404.00,394.27)(6.500,1.000){2}{\rule{1.566pt}{0.350pt}}
\put(417,395.77){\rule{2.891pt}{0.350pt}}
\multiput(417.00,395.27)(6.000,1.000){2}{\rule{1.445pt}{0.350pt}}
\put(429,397.27){\rule{2.188pt}{0.350pt}}
\multiput(429.00,396.27)(7.460,2.000){2}{\rule{1.094pt}{0.350pt}}
\put(441,399.27){\rule{2.188pt}{0.350pt}}
\multiput(441.00,398.27)(7.460,2.000){2}{\rule{1.094pt}{0.350pt}}
\multiput(453.00,401.47)(3.170,0.516){3}{\rule{1.604pt}{0.124pt}}
\multiput(453.00,400.27)(9.670,3.000){2}{\rule{0.802pt}{0.350pt}}
\put(466,404.27){\rule{2.188pt}{0.350pt}}
\multiput(466.00,403.27)(7.460,2.000){2}{\rule{1.094pt}{0.350pt}}
\multiput(478.00,406.47)(2.913,0.516){3}{\rule{1.487pt}{0.124pt}}
\multiput(478.00,405.27)(8.913,3.000){2}{\rule{0.744pt}{0.350pt}}
\multiput(490.00,409.47)(3.170,0.516){3}{\rule{1.604pt}{0.124pt}}
\multiput(490.00,408.27)(9.670,3.000){2}{\rule{0.802pt}{0.350pt}}
\put(503,412.27){\rule{2.188pt}{0.350pt}}
\multiput(503.00,411.27)(7.460,2.000){2}{\rule{1.094pt}{0.350pt}}
\multiput(515.00,414.47)(2.913,0.516){3}{\rule{1.487pt}{0.124pt}}
\multiput(515.00,413.27)(8.913,3.000){2}{\rule{0.744pt}{0.350pt}}
\multiput(527.00,417.47)(2.913,0.516){3}{\rule{1.487pt}{0.124pt}}
\multiput(527.00,416.27)(8.913,3.000){2}{\rule{0.744pt}{0.350pt}}
\multiput(539.00,420.47)(3.170,0.516){3}{\rule{1.604pt}{0.124pt}}
\multiput(539.00,419.27)(9.670,3.000){2}{\rule{0.802pt}{0.350pt}}
\multiput(552.00,423.47)(2.913,0.516){3}{\rule{1.487pt}{0.124pt}}
\multiput(552.00,422.27)(8.913,3.000){2}{\rule{0.744pt}{0.350pt}}
\put(564,426.27){\rule{2.188pt}{0.350pt}}
\multiput(564.00,425.27)(7.460,2.000){2}{\rule{1.094pt}{0.350pt}}
\multiput(576.00,428.47)(2.913,0.516){3}{\rule{1.487pt}{0.124pt}}
\multiput(576.00,427.27)(8.913,3.000){2}{\rule{0.744pt}{0.350pt}}
\multiput(588.00,431.47)(3.170,0.516){3}{\rule{1.604pt}{0.124pt}}
\multiput(588.00,430.27)(9.670,3.000){2}{\rule{0.802pt}{0.350pt}}
\put(601,434.27){\rule{2.188pt}{0.350pt}}
\multiput(601.00,433.27)(7.460,2.000){2}{\rule{1.094pt}{0.350pt}}
\multiput(613.00,436.47)(2.913,0.516){3}{\rule{1.487pt}{0.124pt}}
\multiput(613.00,435.27)(8.913,3.000){2}{\rule{0.744pt}{0.350pt}}
\put(625,439.27){\rule{2.362pt}{0.350pt}}
\multiput(625.00,438.27)(8.097,2.000){2}{\rule{1.181pt}{0.350pt}}
\multiput(638.00,441.47)(2.913,0.516){3}{\rule{1.487pt}{0.124pt}}
\multiput(638.00,440.27)(8.913,3.000){2}{\rule{0.744pt}{0.350pt}}
\put(650,444.27){\rule{2.188pt}{0.350pt}}
\multiput(650.00,443.27)(7.460,2.000){2}{\rule{1.094pt}{0.350pt}}
\multiput(662.00,446.47)(2.913,0.516){3}{\rule{1.487pt}{0.124pt}}
\multiput(662.00,445.27)(8.913,3.000){2}{\rule{0.744pt}{0.350pt}}
\put(674,449.27){\rule{2.362pt}{0.350pt}}
\multiput(674.00,448.27)(8.097,2.000){2}{\rule{1.181pt}{0.350pt}}
\put(687,451.27){\rule{2.188pt}{0.350pt}}
\multiput(687.00,450.27)(7.460,2.000){2}{\rule{1.094pt}{0.350pt}}
\put(699,453.27){\rule{2.188pt}{0.350pt}}
\multiput(699.00,452.27)(7.460,2.000){2}{\rule{1.094pt}{0.350pt}}
\multiput(711.00,455.47)(3.170,0.516){3}{\rule{1.604pt}{0.124pt}}
\multiput(711.00,454.27)(9.670,3.000){2}{\rule{0.802pt}{0.350pt}}
\put(724,458.27){\rule{2.188pt}{0.350pt}}
\multiput(724.00,457.27)(7.460,2.000){2}{\rule{1.094pt}{0.350pt}}
\put(736,460.27){\rule{2.188pt}{0.350pt}}
\multiput(736.00,459.27)(7.460,2.000){2}{\rule{1.094pt}{0.350pt}}
\put(748,462.27){\rule{2.188pt}{0.350pt}}
\multiput(748.00,461.27)(7.460,2.000){2}{\rule{1.094pt}{0.350pt}}
\put(760,463.77){\rule{3.132pt}{0.350pt}}
\multiput(760.00,463.27)(6.500,1.000){2}{\rule{1.566pt}{0.350pt}}
\put(773,465.27){\rule{2.188pt}{0.350pt}}
\multiput(773.00,464.27)(7.460,2.000){2}{\rule{1.094pt}{0.350pt}}
\put(785,467.27){\rule{2.188pt}{0.350pt}}
\multiput(785.00,466.27)(7.460,2.000){2}{\rule{1.094pt}{0.350pt}}
\put(797,469.27){\rule{2.362pt}{0.350pt}}
\multiput(797.00,468.27)(8.097,2.000){2}{\rule{1.181pt}{0.350pt}}
\put(810,470.77){\rule{2.891pt}{0.350pt}}
\multiput(810.00,470.27)(6.000,1.000){2}{\rule{1.445pt}{0.350pt}}
\put(822,472.27){\rule{2.188pt}{0.350pt}}
\multiput(822.00,471.27)(7.460,2.000){2}{\rule{1.094pt}{0.350pt}}
\put(834,474.27){\rule{2.188pt}{0.350pt}}
\multiput(834.00,473.27)(7.460,2.000){2}{\rule{1.094pt}{0.350pt}}
\put(846,475.77){\rule{3.132pt}{0.350pt}}
\multiput(846.00,475.27)(6.500,1.000){2}{\rule{1.566pt}{0.350pt}}
\put(859,477.27){\rule{2.188pt}{0.350pt}}
\multiput(859.00,476.27)(7.460,2.000){2}{\rule{1.094pt}{0.350pt}}
\put(871,478.77){\rule{2.891pt}{0.350pt}}
\multiput(871.00,478.27)(6.000,1.000){2}{\rule{1.445pt}{0.350pt}}
\put(883,480.27){\rule{2.362pt}{0.350pt}}
\multiput(883.00,479.27)(8.097,2.000){2}{\rule{1.181pt}{0.350pt}}
\put(896,481.77){\rule{2.891pt}{0.350pt}}
\multiput(896.00,481.27)(6.000,1.000){2}{\rule{1.445pt}{0.350pt}}
\put(908,482.77){\rule{2.891pt}{0.350pt}}
\multiput(908.00,482.27)(6.000,1.000){2}{\rule{1.445pt}{0.350pt}}
\put(920,484.27){\rule{2.188pt}{0.350pt}}
\multiput(920.00,483.27)(7.460,2.000){2}{\rule{1.094pt}{0.350pt}}
\put(932,485.77){\rule{3.132pt}{0.350pt}}
\multiput(932.00,485.27)(6.500,1.000){2}{\rule{1.566pt}{0.350pt}}
\put(945,486.77){\rule{2.891pt}{0.350pt}}
\multiput(945.00,486.27)(6.000,1.000){2}{\rule{1.445pt}{0.350pt}}
\put(957,487.77){\rule{2.891pt}{0.350pt}}
\multiput(957.00,487.27)(6.000,1.000){2}{\rule{1.445pt}{0.350pt}}
\put(969,489.27){\rule{2.362pt}{0.350pt}}
\multiput(969.00,488.27)(8.097,2.000){2}{\rule{1.181pt}{0.350pt}}
\put(982,490.77){\rule{2.891pt}{0.350pt}}
\multiput(982.00,490.27)(6.000,1.000){2}{\rule{1.445pt}{0.350pt}}
\put(994,491.77){\rule{2.891pt}{0.350pt}}
\multiput(994.00,491.27)(6.000,1.000){2}{\rule{1.445pt}{0.350pt}}
\put(1006,492.77){\rule{2.891pt}{0.350pt}}
\multiput(1006.00,492.27)(6.000,1.000){2}{\rule{1.445pt}{0.350pt}}
\put(1018,493.77){\rule{3.132pt}{0.350pt}}
\multiput(1018.00,493.27)(6.500,1.000){2}{\rule{1.566pt}{0.350pt}}
\put(1031,494.77){\rule{2.891pt}{0.350pt}}
\multiput(1031.00,494.27)(6.000,1.000){2}{\rule{1.445pt}{0.350pt}}
\put(1043,495.77){\rule{2.891pt}{0.350pt}}
\multiput(1043.00,495.27)(6.000,1.000){2}{\rule{1.445pt}{0.350pt}}
\put(1055,496.77){\rule{3.132pt}{0.350pt}}
\multiput(1055.00,496.27)(6.500,1.000){2}{\rule{1.566pt}{0.350pt}}
\put(1068,497.77){\rule{2.891pt}{0.350pt}}
\multiput(1068.00,497.27)(6.000,1.000){2}{\rule{1.445pt}{0.350pt}}
\put(1080,498.77){\rule{2.891pt}{0.350pt}}
\multiput(1080.00,498.27)(6.000,1.000){2}{\rule{1.445pt}{0.350pt}}
\put(1092,499.77){\rule{2.891pt}{0.350pt}}
\multiput(1092.00,499.27)(6.000,1.000){2}{\rule{1.445pt}{0.350pt}}
\put(1104,500.77){\rule{3.132pt}{0.350pt}}
\multiput(1104.00,500.27)(6.500,1.000){2}{\rule{1.566pt}{0.350pt}}
\put(1129,501.77){\rule{2.891pt}{0.350pt}}
\multiput(1129.00,501.27)(6.000,1.000){2}{\rule{1.445pt}{0.350pt}}
\put(1141,502.77){\rule{2.891pt}{0.350pt}}
\multiput(1141.00,502.27)(6.000,1.000){2}{\rule{1.445pt}{0.350pt}}
\put(1153,503.77){\rule{3.132pt}{0.350pt}}
\multiput(1153.00,503.27)(6.500,1.000){2}{\rule{1.566pt}{0.350pt}}
\put(1166,504.77){\rule{2.891pt}{0.350pt}}
\multiput(1166.00,504.27)(6.000,1.000){2}{\rule{1.445pt}{0.350pt}}
\put(1117.0,502.0){\rule[-0.175pt]{2.891pt}{0.350pt}}
\put(1190,505.77){\rule{3.132pt}{0.350pt}}
\multiput(1190.00,505.27)(6.500,1.000){2}{\rule{1.566pt}{0.350pt}}
\put(1203,506.77){\rule{2.891pt}{0.350pt}}
\multiput(1203.00,506.27)(6.000,1.000){2}{\rule{1.445pt}{0.350pt}}
\put(1215,507.77){\rule{2.891pt}{0.350pt}}
\multiput(1215.00,507.27)(6.000,1.000){2}{\rule{1.445pt}{0.350pt}}
\put(1178.0,506.0){\rule[-0.175pt]{2.891pt}{0.350pt}}
\put(1239,508.77){\rule{3.132pt}{0.350pt}}
\multiput(1239.00,508.27)(6.500,1.000){2}{\rule{1.566pt}{0.350pt}}
\put(1227.0,509.0){\rule[-0.175pt]{2.891pt}{0.350pt}}
\put(1264,509.77){\rule{2.891pt}{0.350pt}}
\multiput(1264.00,509.27)(6.000,1.000){2}{\rule{1.445pt}{0.350pt}}
\put(1276,510.77){\rule{3.132pt}{0.350pt}}
\multiput(1276.00,510.27)(6.500,1.000){2}{\rule{1.566pt}{0.350pt}}
\put(1252.0,510.0){\rule[-0.175pt]{2.891pt}{0.350pt}}
\put(1301,511.77){\rule{2.891pt}{0.350pt}}
\multiput(1301.00,511.27)(6.000,1.000){2}{\rule{1.445pt}{0.350pt}}
\put(1313,512.77){\rule{2.891pt}{0.350pt}}
\multiput(1313.00,512.27)(6.000,1.000){2}{\rule{1.445pt}{0.350pt}}
\put(1289.0,512.0){\rule[-0.175pt]{2.891pt}{0.350pt}}
\put(1338,513.77){\rule{2.891pt}{0.350pt}}
\multiput(1338.00,513.27)(6.000,1.000){2}{\rule{1.445pt}{0.350pt}}
\put(1325.0,514.0){\rule[-0.175pt]{3.132pt}{0.350pt}}
\put(1362,514.77){\rule{3.132pt}{0.350pt}}
\multiput(1362.00,514.27)(6.500,1.000){2}{\rule{1.566pt}{0.350pt}}
\put(1350.0,515.0){\rule[-0.175pt]{2.891pt}{0.350pt}}
\put(1387,515.77){\rule{2.891pt}{0.350pt}}
\multiput(1387.00,515.27)(6.000,1.000){2}{\rule{1.445pt}{0.350pt}}
\put(1375.0,516.0){\rule[-0.175pt]{2.891pt}{0.350pt}}
\put(1411,516.77){\rule{3.132pt}{0.350pt}}
\multiput(1411.00,516.27)(6.500,1.000){2}{\rule{1.566pt}{0.350pt}}
\put(1399.0,517.0){\rule[-0.175pt]{2.891pt}{0.350pt}}
\put(1424.0,518.0){\rule[-0.175pt]{2.891pt}{0.350pt}}
\sbox{\plotpoint}{\rule[-0.250pt]{0.500pt}{0.500pt}}%
\put(555,612){\makebox(0,0)[r]{$10^7\Delta \sigma_{yy}$}}
\multiput(577,612)(12.453,0.000){6}{\usebox{\plotpoint}}
\put(643,612){\usebox{\plotpoint}}
\put(220,771){\usebox{\plotpoint}}
\multiput(220,771)(0.765,-12.430){16}{\usebox{\plotpoint}}
\multiput(232,576)(0.996,-12.413){13}{\usebox{\plotpoint}}
\multiput(245,414)(1.414,-12.373){9}{\usebox{\plotpoint}}
\multiput(257,309)(2.482,-12.203){5}{\usebox{\plotpoint}}
\multiput(269,250)(4.762,-11.507){2}{\usebox{\plotpoint}}
\put(285.30,218.02){\usebox{\plotpoint}}
\put(295.78,212.59){\usebox{\plotpoint}}
\multiput(306,216)(9.180,8.415){2}{\usebox{\plotpoint}}
\put(324.26,235.19){\usebox{\plotpoint}}
\multiput(331,244)(6.650,10.529){2}{\usebox{\plotpoint}}
\multiput(343,263)(6.407,10.679){2}{\usebox{\plotpoint}}
\multiput(355,283)(6.407,10.679){2}{\usebox{\plotpoint}}
\multiput(367,303)(6.787,10.441){2}{\usebox{\plotpoint}}
\multiput(380,323)(6.650,10.529){2}{\usebox{\plotpoint}}
\put(397.79,350.69){\usebox{\plotpoint}}
\multiput(404,360)(7.565,9.892){2}{\usebox{\plotpoint}}
\multiput(417,377)(7.472,9.963){2}{\usebox{\plotpoint}}
\put(435.05,400.56){\usebox{\plotpoint}}
\multiput(441,408)(8.105,9.455){2}{\usebox{\plotpoint}}
\put(459.90,428.37){\usebox{\plotpoint}}
\multiput(466,434)(8.806,8.806){2}{\usebox{\plotpoint}}
\put(486.91,454.17){\usebox{\plotpoint}}
\put(496.55,462.04){\usebox{\plotpoint}}
\put(506.45,469.59){\usebox{\plotpoint}}
\multiput(515,476)(10.362,6.908){2}{\usebox{\plotpoint}}
\put(537.19,490.80){\usebox{\plotpoint}}
\put(548.05,496.87){\usebox{\plotpoint}}
\put(559.13,502.56){\usebox{\plotpoint}}
\put(570.27,508.13){\usebox{\plotpoint}}
\put(581.41,513.70){\usebox{\plotpoint}}
\put(592.86,518.49){\usebox{\plotpoint}}
\put(604.63,522.51){\usebox{\plotpoint}}
\put(616.21,527.07){\usebox{\plotpoint}}
\put(628.05,530.94){\usebox{\plotpoint}}
\put(639.98,534.50){\usebox{\plotpoint}}
\put(652.02,537.67){\usebox{\plotpoint}}
\put(663.90,541.32){\usebox{\plotpoint}}
\put(676.16,543.50){\usebox{\plotpoint}}
\put(688.29,546.32){\usebox{\plotpoint}}
\put(700.40,549.23){\usebox{\plotpoint}}
\put(712.68,551.26){\usebox{\plotpoint}}
\put(724.99,553.16){\usebox{\plotpoint}}
\put(737.29,555.11){\usebox{\plotpoint}}
\put(749.68,556.28){\usebox{\plotpoint}}
\put(761.98,558.15){\usebox{\plotpoint}}
\put(774.39,559.23){\usebox{\plotpoint}}
\put(786.69,561.14){\usebox{\plotpoint}}
\put(799.10,562.16){\usebox{\plotpoint}}
\put(811.51,563.13){\usebox{\plotpoint}}
\put(823.92,564.16){\usebox{\plotpoint}}
\put(836.33,565.19){\usebox{\plotpoint}}
\put(848.75,566.00){\usebox{\plotpoint}}
\put(861.20,566.18){\usebox{\plotpoint}}
\put(873.61,567.22){\usebox{\plotpoint}}
\put(886.03,568.00){\usebox{\plotpoint}}
\put(898.48,568.21){\usebox{\plotpoint}}
\put(910.90,569.00){\usebox{\plotpoint}}
\put(923.34,569.28){\usebox{\plotpoint}}
\put(935.76,570.00){\usebox{\plotpoint}}
\put(948.21,570.00){\usebox{\plotpoint}}
\put(960.67,570.00){\usebox{\plotpoint}}
\put(973.11,570.32){\usebox{\plotpoint}}
\put(985.54,571.00){\usebox{\plotpoint}}
\put(997.99,571.00){\usebox{\plotpoint}}
\put(1010.44,571.00){\usebox{\plotpoint}}
\put(1022.90,571.00){\usebox{\plotpoint}}
\put(1035.35,571.00){\usebox{\plotpoint}}
\put(1047.79,571.40){\usebox{\plotpoint}}
\put(1060.21,572.00){\usebox{\plotpoint}}
\put(1072.67,572.00){\usebox{\plotpoint}}
\put(1085.12,572.00){\usebox{\plotpoint}}
\put(1097.57,572.00){\usebox{\plotpoint}}
\put(1110.03,572.00){\usebox{\plotpoint}}
\put(1122.48,572.00){\usebox{\plotpoint}}
\put(1134.93,572.00){\usebox{\plotpoint}}
\put(1147.39,572.00){\usebox{\plotpoint}}
\put(1159.82,571.48){\usebox{\plotpoint}}
\put(1172.25,571.00){\usebox{\plotpoint}}
\put(1184.71,571.00){\usebox{\plotpoint}}
\put(1197.16,571.00){\usebox{\plotpoint}}
\put(1209.61,571.00){\usebox{\plotpoint}}
\put(1222.07,571.00){\usebox{\plotpoint}}
\put(1234.52,571.00){\usebox{\plotpoint}}
\put(1246.97,571.00){\usebox{\plotpoint}}
\put(1259.43,571.00){\usebox{\plotpoint}}
\put(1271.88,571.00){\usebox{\plotpoint}}
\put(1284.33,571.00){\usebox{\plotpoint}}
\put(1296.76,570.35){\usebox{\plotpoint}}
\put(1309.20,570.00){\usebox{\plotpoint}}
\put(1321.65,570.00){\usebox{\plotpoint}}
\put(1334.11,570.00){\usebox{\plotpoint}}
\put(1346.56,570.00){\usebox{\plotpoint}}
\put(1359.01,570.00){\usebox{\plotpoint}}
\put(1371.47,570.00){\usebox{\plotpoint}}
\put(1383.92,570.00){\usebox{\plotpoint}}
\put(1396.34,569.22){\usebox{\plotpoint}}
\put(1408.78,569.00){\usebox{\plotpoint}}
\put(1421.24,569.00){\usebox{\plotpoint}}
\put(1433.69,569.00){\usebox{\plotpoint}}
\put(1436,569){\usebox{\plotpoint}}
\end{picture}}
\end{picture}
\caption[]{\label{boltz1} The calculated zero-temperature conductivity
$\sigma _{0}$ (upper curve) as a function of
the Fermi energy $\varepsilon _{F}$ in a 2d Silicon accumulation layer.
The scattering of electrons is due to impurities.
A microwave field $E_{0}$ (linearly polarized in $x$-direction) with
frequency $\Omega =10^{10} s^{-1}$ gives rise to changes $\Delta \sigma
_{xx}$ and $\Delta \sigma _{yy}$ of the conductivity
(note the different scales), calculated to second
order in $E_{0}$ according to Eq.~(\ref{finalsecond}). Parameters
for the relaxation time as a function of $\varepsilon _{F}$ due to impurity
scattering were chosen according to \cite{IS86}. Curves are shown for
$E_{0}=1 V m^{-1}$ and have to be scaled with $E_{0}^{2}$ for other values
of the microwave field.}
\end{figure}

Indeed, in realistic systems this dependence can be quite strong.
Years ago, Fang, Fowler and Hartstein
\cite{Fanetal77} measured the Shubnikov-de-Haas oscillations in a Si
inversion layer as a function of the gate voltage. The general shape of the
magnetoconductivity  curve was determined by a broad maximum when plotted
against the Fermi energy (compare Fig. \ref{boltz1}). This feature was
independent of the magnetic field $B$ and could, beside
the $B$--dependent oscillations, be reproduced
in a theoretical calculation by Isihara and Smr\v{c}ka \cite{IS86}. They used a
pseudopotential model for impurity scattering
and a simple CPA approach
corresponding to the relaxation time approximation in the Boltzmann equation.
The key point was the functional dependence of $\tau
(\varepsilon )$, the scattering time, on the energy, namely
\begin{equation}
\label{tauepsilon}
\tau (\varepsilon )=a\left[\frac{\varepsilon _{\sigma }\varepsilon }
{\varepsilon _{\sigma }\varepsilon ^{2}+\varepsilon (\varepsilon _{\sigma
}^{2}-3\varepsilon _{\tau }^{2})+\varepsilon _{\sigma }\varepsilon _{\tau
}^{2}}\right]^{2},
\end{equation}
in lowest order in the impurity
concentration. Parameters fitting the experiment \cite{Fanetal77} where
obtained as $a=4.3*10^{-10}$ meV$^{2}$s, $\varepsilon _{\tau }=3.5$ meV and
$\varepsilon _{\sigma }=15$ meV.
We used Eq.~(\ref{tauepsilon}) as an example for the evaluation of
Eq.~(\ref{finalsecond}). The result is shown in the Fig. (\ref{boltz1}).

An important feature is that under the influence of the microwave the
conductivity tensor becomes non-isotropic, i.e. $\sigma _{xx}$ and $\sigma
_{yy}$ are different for $E_{0}\ne 0$ (remember that the microwave is assumed
to be polarized in $x$-direction). Moreover, the signs of
the change $\Delta \sigma
_{xx}$ and $\Delta \sigma _{yy}$ due to the microwave depend on the value of
the Fermi energy
$\varepsilon _{F}$. Indeed, the different terms in Eq.~(\ref{finalsecond})
involving both first and second derivative of $\tau ^{-1}(\varepsilon _{F})$
can even cancel for a certain value of $\varepsilon _{F}$ as is seen in the
curve for $\Delta \sigma _{yy}$.

All these features should be detectable in an experiment. For direct
comparison with the theoretical result, it would be preferable to work with
linearly polarized microwaves with wavelengths longer than any relevant
microscopic length scale in the system so that the microwave can be treated
in the ${\bf q}=0$--limit. Furthermore, other materials then silicon could be
used as well, in this case the model for the scattering time
Eq.~(\ref{tauepsilon}) of course has to
be modified. However, as long as there is a noticeable dependence of the
scattering rate on the energy, the microwave will change the conductance
according to Eq.~(\ref{finalsecond}).
On the other hand, one could imagine a system where the dependence of the
scattering rate on the Fermi energy is unknown and not as easily experimentally
detectable as was the case in \cite{Fanetal77}. Then, the microwave
could serve as a probe to extract this dependence via
Eq.~(\ref{finalsecond}).

Since we calculated only to second
order in $E_{0}$, the applicability of Eq.~(\ref{finalsecond}) is restricted
to not too high microwave field strengths. Indeed, the numerical evaluation
Fig. (\ref{boltz1}) shows that at $E_{0}\approx 1000 V/m$ the change $\Delta
\sigma _{xx}$, $\Delta \sigma _{yy}$ becomes of the order of the conductivity
$\sigma _{0}$ itself, indicating that above this value of $E_{0}$ one has to
go at least to fourth order perturbation theory in the microwave.
Furthermore, we did not consider effects like electron heating, ionization or
Zener breakdown due to the microwave field but purely the interplay between
the momentum conserving impurity scattering  and the kinematics of electrons
in an oscillating field.

As a summary, we presented a theoretical approach to linear transport in
the presence of microwave irradiation within a semiclassical description. The
general result for the conductivity was discussed for the case of a two-
dimensional Silicon inversion layer, and deviations from the Drude result were
predicted to be observable in the zero magnetic field case.

The author would like to acknowledge discussions with A. Kawabata, B. Kramer,
G. Platero, R. Kilian and V. Reinstorf, and financial support by the EU STF9
fellowship program in Japan.

\end{document}